# Evolutionary tuning of TAM receptor–ligand interfaces highlights electrostatic features associated with regenerative phagocytic signaling

Enso O. Torres Alegre

ORCID: 0000-0002-6798-8776

Pontifical Catholic University of Chile, Santiago, Chile

onill@uc.cl

## Abstract

Efficient resolution of neuroinflammation and debris clearance are key determinants of successful central nervous system (CNS) regeneration. Regenerative vertebrates such as *Danio rerio* often show faster immune resolution and debris clearance than mammals, yet the molecular determinants underlying these differences remain incompletely understood. TAM receptor tyrosine kinases (Tyro3, Axl, and Mertk) and their ligands Gas6 and Protein S are central regulators of phagocytosis and immune resolution in the nervous system, but whether intrinsic structural properties of these receptor–ligand complexes contribute to regenerative efficiency has not been systematically explored.

Here, we perform a comparative in silico analysis of TAM receptors and ligands from zebrafish, human, and mouse, integrating sequence evolution, high-confidence structural modeling, interface characterization, and electrostatic analysis. Despite substantial sequence divergence between mammals and zebrafish, ligand-binding domains retain strong structural conservation, supporting a conserved global mode of TAM–ligand engagement. At the interface level, zebrafish complexes exhibit enhanced electrostatic contributions and increased salt-bridge density, particularly in the Tyro3–Protein S interaction. Residue-resolved electrostatic analysis identifies clustered interface hotspots that are conserved in spatial organization and physicochemical function across species, despite evolutionary rewiring of individual contacts.

Together, these findings suggest that TAM receptor–ligand interfaces are evolutionarily tuned through subtle electrostatic and geometric optimization rather than large-scale structural changes. This conserved yet adaptable electrostatic framework supports a mechanistic hypothesis in which conserved TAM receptor architecture permits species-specific electrostatic tuning of receptor–ligand interfaces, potentially contributing to differences in TAM-dependent phagocytic signaling efficiency across vertebrates

## 1. Introduction

Efficient regeneration of the adult central nervous system (CNS) remains one of the major challenges in neuroscience. In mammals, traumatic injury or neurodegenerative processes typically elicit prolonged neuroinflammatory responses, inefficient clearance of cellular debris, and the formation of an inhibitory microenvironment that severely restricts functional recovery [1], [2], [3]. In contrast, regenerative vertebrates such as the zebrafish (Danio rerio) display a remarkable capacity for CNS repair, including robust axonal regrowth, restoration of neural circuitry, and functional recovery after spinal cord injury. Comparative studies have consistently demonstrated that this regenerative competence is closely associated with rapid immune resolution and highly efficient phagocytic clearance of apoptotic cells, myelin debris, and damaged cellular debris [4], [5], [6], [7].

Phagocytosis in the injured nervous system is not a passive waste-disposal process but a highly regulated and instructive mechanism that actively shapes tissue repair [8][9]. Microglia, macrophages, astrocytes, and peripheral glial cells dynamically coordinate the recognition, engulfment, and processing of debris while simultaneously modulating inflammatory signaling, extracellular matrix remodeling, synaptogenesis, and neurogenesis. Dysregulated phagocytic activity can exacerbate tissue damage through excessive inflammation or inappropriate engulfment of stressed but viable cells, whereas insufficient clearance leads to chronic immune activation and regenerative failure [4], [8]. Thus, the molecular pathways that fine-tune phagocytosis and efferocytosis represent critical control points for balancing inflammation and repair in the CNS. A key open question is whether part of this control is encoded not only at the level of expression and downstream signaling, but also in the intrinsic biophysical properties of receptor–ligand interfaces that initiate phagocytic programs.

Among the molecular regulators of phagocytosis, the TAM family of receptor tyrosine kinases—Tyro3, Axl, and Mertk—has emerged as a central signaling hub. Activated by the vitamin-K-dependent ligands Growth Arrest-Specific 6 (Gas6) and Protein S (Pros1), TAM receptors mediate phosphatidylserine-dependent recognition of apoptotic cells [3] and orchestrate anti-inflammatory and pro-repair responses across multiple tissues, including the nervous system [10], [11], [12]. In neural contexts, TAM signaling regulates microglial activation, astrocytic and Schwann-cell-mediated debris clearance [13], myelination, synaptic remodeling, neuronal survival, and immune homeostasis [3], [8]. Disruption of TAM signaling leads to impaired phagocytosis, sustained inflammation, and contributes to neuroinflammatory and neurodegenerative disorders such as Alzheimer's disease, Parkinson's disease, multiple sclerosis, traumatic brain injury, and spinal cord injury [8], [10].

A growing body of experimental evidence indicates that enhancing TAM signaling improves outcomes after CNS injury. Activation of the Gas6–Axl or Gas6–Mertk axes promotes microglial and astrocytic phagocytosis, suppresses pro-inflammatory cytokine production, stabilizes vascular and barrier integrity, and facilitates neural repair following spinal cord injury, traumatic brain injury, experimental autoimmune encephalomyelitis, and sepsis-associated encephalopathy [14], [15], [16], [17], [18], [19]. Conversely, genetic ablation or pharmacological inhibition of TAM receptors impairs debris clearance and exacerbates inflammatory pathology. Together, these studies establish TAM receptors as key effectors of repair-oriented immune responses in the nervous system and highlight their translational relevance as therapeutic targets.

Within the TAM family, receptor–ligand interactions operate across distinct functional regimes. Mertk–Gas6 represents a canonical efferocytosis axis tightly linked to homeostatic apoptotic cell clearance, whereas Axl–Gas6 is preferentially engaged in inflammatory and injury-associated phagocytic responses. In contrast, the Tyro3–Protein S interaction remains comparatively less characterized and appears to operate under looser evolutionary constraint, potentially allowing greater structural and electrostatic plasticity.

Despite this extensive functional and translational evidence, a fundamental question remains unresolved: **why is TAM signaling more effective in regenerative contexts such as the zebrafish CNS than in mammals?** Current models primarily emphasize transcriptional regulation, ligand availability, or downstream signaling pathways. However, the possibility that intrinsic, species-specific structural properties of TAM receptors and their ligand-binding interfaces contribute to differences in signaling efficiency has not been systematically explored. Whether variations in receptor–ligand affinity, interface geometry, or complex stability underlie enhanced phagocytic signaling in regenerative species remains unknown.

Importantly, recent structural and biophysical studies in zebrafish have shown that immune-related proteins unique to teleosts can preserve conserved structural organization and ligand-dependent activation mechanisms comparable to their mammalian counterparts, despite substantial evolutionary divergence [20].

Recent advances in protein structure prediction and computational biophysics now enable comparative, atomistic analyses of receptor–ligand complexes across species with unprecedented accuracy. Deep-learning-based methods such as AlphaFold2 and AlphaFold-Multimer have revolutionized high-confidence structural modeling of protein complexes [21], [22], [23], while tools for interface analysis and binding-energy estimation allow quantitative assessment of molecular interactions [24], [25].

Integrating these approaches with evolutionary analysis provides a powerful framework to identify structural determinants that may have been selectively optimized in regenerative lineages.

In this study, we perform a comprehensive in silico comparative analysis of TAM receptors (Tyro3, Axl, and Mertk) and their ligands Gas6 and Protein S in zebrafish and mammals. By combining evolutionary sequence analysis, high-confidence structural modeling, interface characterization, binding-energy estimation, and residue-resolved electrostatic analysis, we identify species-specific features of TAM–ligand interfaces that may contribute to enhanced signaling efficiency in regenerative contexts. Our results provide an integrated structural and evolutionary framework linking TAM receptor–ligand interactions to differential regenerative capacity and establish a rational basis for future experimental and computational studies aimed at modulating TAM signaling in the mammalian nervous system.

## 2. Materials and Methods

### 2.1 Sequence Retrieval and Orthology Definition

Protein sequences of the TAM receptors (Tyro3, Axl, and Mertk) and their ligands Growth Arrest–Specific 6 (Gas6) and Protein S (Pros1) were retrieved in FASTA format from UniProtKB for Danio rerio, Homo sapiens, and Mus musculus. Only reviewed or canonical isoforms were selected to ensure sequence consistency across species. Orthology assignments were verified using UniProt annotations and reciprocal sequence similarity.

### 2.2 Multiple Sequence Alignment and Sequence Identity Analysis

Multiple sequence alignments were performed using MAFFT (v7) with default parameters, a tool optimized for accurate alignment of homologous protein sequences [26]. Pairwise global sequence identities were calculated using the EMBOSS Needle program, which implements the Needleman–Wunsch global alignment algorithm [27], executed via the EMBL-EBI web-based Job Dispatcher framework [28].

### 2.3 Pairwise Sequence Identity Analysis

Pairwise global sequence identities were calculated for full-length TAM receptors and ligands, as well as for individual ligand-binding domains following structural segmentation. Global alignments were performed using the EMBOSS Needle program, which implements the Needleman–Wunsch algorithm. Pairwise identity matrices were generated and visualized as heatmaps to highlight patterns of conservation and divergence between species.

### 2.3.1 Structure Prediction of Individual TAM Receptors and Ligands

Structural models of TAM receptors and ligands from each species were generated using AlphaFold2, implemented via ColabFold, which enables efficient and high-confidence structure prediction [21], [23]. Default ColabFold parameters were used unless otherwise specified.

Model quality was assessed using per-residue confidence scores (pLDDT). Only models in which the extracellular ligand-binding domains exhibited consistently high confidence (pLDDT > 80 for the majority of residues) were retained for subsequent analyses.

### 2.4 Domain Extraction and Structural Alignment

To focus on receptor–ligand interaction regions, predicted structures were imported into UCSF ChimeraX [29] for domain segmentation and visualization. The following extracellular domains were extracted based on known TAM receptor architecture:

- Receptors: Ig-like domains 1 and 2
- Ligands: LG domains 1 and 2

Extracted domains were structurally aligned using UCSF ChimeraX [29], and root-mean-square deviation (RMSD) values were calculated to quantify structural similarity across species.

Additionally, structural similarity was independently assessed using TM-align (Zhang Lab), which provides a length-independent metric for fold comparison [30]. TM-scores were used to complement RMSD measurements and ensure robustness of structural comparisons.

### 2.5 Comparative Analysis of Sequence and Structural Conservation

Sequence identities of both full-length proteins and extracted interaction domains were compared across species to assess evolutionary conservation at different structural levels. Heatmaps of sequence identity were generated to visualize conservation patterns among receptors and ligands.

Structural similarity metrics (RMSD and TM-score) were integrated with sequence identity data to evaluate whether sequence divergence translated into measurable structural differences within ligand-binding regions.

### 2.6 Modeling of TAM–Ligand Interaction Complexes

Protein–protein interaction models were generated for the following biologically relevant receptor–ligand pairs:

- Gas6–Mertk

- Gas6–Axl

- Protein S–Tyro3

To specifically focus on the ligand-binding interfaces, only the extracellular receptor-binding domains (RBDs) were modeled rather than full-length proteins. For TAM receptors, this corresponded to the first and second immunoglobulin-like domains (Ig-like 1 and Ig-like 2), which mediate ligand recognition. For the ligands Gas6 and Protein S, the modeled regions comprised the first and second laminin G–like domains (LG1 and LG2), which are known to interact with TAM receptors.

Receptor and ligand RBD sequences were extracted from the full-length proteins and used as input for AlphaFold-Multimer, as implemented in ColabFold, to predict receptor–ligand assemblies directly from amino acid sequences [21], [22], [23]. For each receptor–ligand pair, interaction models were generated independently for zebrafish (Danio rerio), human (Homo sapiens), and mouse (Mus musculus) orthologs.

Model quality was assessed using AlphaFold-Multimer confidence outputs (ipTM, pTM, per-chain pLDDT, interface pLDDT, and mean interface PAE) and interface contact statistics (atom–atom contacts within 4.5 Å). For each species and receptor–ligand pair, the highest-confidence models were selected for subsequent interface and comparative analyses; summary metrics are reported in Supplementary Table S4

## 2.7 Interface Characterization and Binding Analysis

Protein–protein interfaces of the predicted complexes were analyzed using PDBePISA, which identifies biologically relevant interfaces and provides quantitative geometric and energetic descriptors [24].

For each complex, the following parameters were extracted:

- Buried surface area (BSA)
- Estimated binding energy
- Hydrogen bonds and salt bridges
- Lists of interacting residues at the receptor–ligand interface

Interface residues were cataloged for downstream comparative and evolutionary analyses.

### 2.7.1 Salt-bridge and hydrogen-bond density calculation

Salt-bridge (NSB) and hydrogen-bond (NHB) densities were calculated to enable size-normalized comparisons of interfacial interaction networks across species and TAM receptor–ligand pairs. For each receptor-binding domain (RBD) complex, NSB, NHB, and buried interface area (Å$^2$) were extracted from PDBePISA analyses (Section 2.7).

Interaction densities were computed as:

$$\rho \text{SB} = \frac{NSB}{A_{Interface}} \times 1000$$

$$\rho \text{HB} = \frac{NHB}{A_{Interface}} \times 1000$$

Where $NSB$ and $NHB$ are the numbers of salt bridges and hydrogen bonds, respectively, and $A_{Interface}$ is the buried interface area in Å$^2$. Densities are reported as the number of interactions per 1000 Å$^2$ of interface area.

This normalization accounts for differences in interface size and allows direct cross-species comparison of electrostatic (salt-bridge) versus polar (hydrogen-bond) interaction packing at TAM–ligand interfaces.

### 2.8 Software and Visualization

All structural visualizations and figure renderings were generated using UCSF ChimeraX [29]. Sequence alignments and identity matrices were processed using custom scripts and standard bioinformatics tools. Heatmaps and comparative plots were generated using standard scientific plotting libraries.

### 2.9 Contact-resolved Interface Electrostatic Analysis (APBS-derived)

Electrostatic properties of receptor–ligand interfaces were analyzed using a custom Python-based workflow implemented in a Jupyter/Google Colab environment. Structural models of receptor–ligand complexes were converted from PDB to PQR format using PDB2PQR (v3.7.1) with the AMBER force field, assigning protonation states at physiological pH (7.4) while preserving chain identifiers.

Electrostatic potentials were computed by solving the linearized Poisson–Boltzmann equation using APBS (Adaptive Poisson–Boltzmann Solver, v3.4.1) [31] with a protein dielectric constant of 2.0, solvent dielectric constant of 78.54, and an ionic strength of 0.15 M. Electrostatic potential maps were generated in OpenDX (.dx) format.

Protein–protein interface residues were defined based on hydrogen bonds and salt bridges identified by PDBePISA (Section 2.7). Interface interaction data were reformatted into simplified CSV files containing receptor residue identity, ligand residue identity, residue indices, interatomic distances, and contact multiplicity.

Electrostatic potential values from APBS grids were mapped onto atomic coordinates of the corresponding PQR structures using trilinear interpolation. For each interface residue, mean electrostatic potentials were computed and subsequently combined across interacting residue pairs to quantify electrostatic potential differences ($\Delta\phi$) and their absolute magnitudes ($|\Delta\phi|$). To account for interaction strength, electrostatic metrics were additionally weighted by the number of atomic contacts per residue pair, yielding contact-weighted descriptors of interface electrostatic complementarity. These quantitative electrostatic descriptors were then used for cross-species comparisons of receptor–ligand interfaces.

## 2.10 Sequence alignment and conservation analysis of electrostatic interface residues

To assess the evolutionary conservation of electrostatic interface residues identified by APBS analysis, a sequence-based comparative analysis was performed between Homo sapiens and Danio rerio. Full-length amino acid sequences of Tyro3 (receptor) and Protein S (ligand) were retrieved from UniProt and analyzed independently.

Multiple sequence alignments were generated separately for receptor and ligand proteins using the MAFFT algorithm, as implemented within the Jalview 2.11.5.1 [32] environment. Global alignment settings were applied to ensure consistent positional correspondence across homologous regions while preserving alignment accuracy in functionally relevant domains.

Residue indices derived from PDBePISA interface analysis, originally reported using truncated structural models, were mapped to full-length sequence numbering using empirically determined residue offsets (Section 2.9). Electrostatic interface residues were defined as those participating in hydrogen bonds or salt bridges and belonging to the top-ranking receptor–ligand residue pairs based on the mean absolute electrostatic potential difference ($|\Delta\phi|$) obtained from APBS calculations.

Interface residues were mapped onto the sequence alignments and visually annotated within Jalview 2.11.5.1 to enable direct comparison between species. Conservation was evaluated based on residue-type similarity and electrostatic role (acidic, basic, polar, or neutral) at aligned positions rather than strict amino acid identity. Electrostatic interactions were classified as directly conserved, functionally conserved, or non-conserved according to the preservation of physicochemical properties and charge complementarity at corresponding interface positions.

### 2.11 Nomenclature

Protein names are reported using standard UniProt conventions (Tyro3, Axl, Mertk, Gas6, and Protein S), independent of species, unless otherwise specified. Species names are italicized at first mention (Homo sapiens, Mus musculus, Danio rerio) and subsequently referred to as human, mouse, and zebrafish. Receptor–ligand complexes are denoted using an en dash (e.g., Tyro3–Protein S). Residue numbering corresponds to full-length protein sequences unless explicitly stated. Throughout the manuscript, species are referred to as human, mouse, and zebrafish in figures and captions, corresponding to Homo sapiens, Mus musculus, and Danio rerio, respectively.

## 3 Results

### 3.1 Sequence and structural conservation of TAM receptors and ligands across species

Pairwise sequence identity analyses revealed high conservation between mammalian orthologs for all TAM receptors and ligands, with human–mouse identities ranging from approximately 79% to 89% across both full-length proteins and ligand-binding domains (Figure 1A,B). In contrast, comparisons between mammalian proteins and zebrafish orthologs showed moderate to low sequence identity, typically ranging from 30% to 50%, consistent with the evolutionary distance between these species.

Domain-focused analyses of the ligand-binding regions (Ig-like domains 1–2 in receptors and LG domains 1–2 in ligands) did not reveal increased sequence conservation relative to full-length proteins. In several cases, ligand-binding domains exhibited equal or lower sequence identity compared to the corresponding full-length sequences, indicating substantial divergence at the primary sequence level within the interaction interfaces (Figure 1B).

Despite this sequence divergence, structural comparisons revealed a markedly different pattern. TM-align analyses demonstrated strong structural conservation of ligand-binding domains across species, with TM-scores typically exceeding 0.8 for receptor Ig-like domains and for the LG domains of Gas6 (Figure 1D). These results indicate preservation of the overall fold and spatial organization of interaction domains, even in the presence of significant sequence divergence. Notably, the LG domains of zebrafish Protein S display substantially lower structural similarity to mammalian counterparts (TM-score ~ 0.46), raising the question of how functional receptor engagement is preserved despite this divergence (Figure 1D).

In contrast, full-length structural comparisons yielded lower and more variable TM-scores (Figure 1C), highlighting the contribution of non-interacting regions to structural variability and supporting a domain-focused strategy for analyzing receptor–ligand interactions.

**Figure 1. Sequence and structural conservation of TAM receptors and ligands across species.**
(A) Pairwise sequence identity heatmaps for full-length TAM receptors (Axl, Mertk, Tyro3) and ligands (Gas6 and Protein S) across human, mouse, and zebrafish orthologs.
(B) Sequence identity heatmaps restricted to ligand-binding domains (Ig-like domains 1–2 for receptors and LG domains 1–2 for ligands).
(C) Structural similarity of full-length proteins assessed by TM-score.
(D) Structural similarity of ligand-binding domains assessed by TM-score, revealing strong fold conservation despite sequence divergence, with Protein S showing marked structural divergence in zebrafish.

### 3.2 Structural modeling of TAM–ligand receptor-binding domain complexes across species

To assess whether sequence divergence among species translates into differences at the level of receptor–ligand assembly, we next examined the structural organization of TAM receptor–ligand receptor-binding domain (RBD) complexes across human, mouse, and zebrafish orthologs. Interaction models were generated using AlphaFold-Multimer, focusing exclusively on the extracellular ligand-binding regions of the receptors (Ig-like domains 1–2) and ligands (LG domains 1–2). Across all modeled complexes, AlphaFold-Multimer confidence metrics and interface contact statistics indicated high structural reliability for comparative interface analysis (ipTM/pTM, per-chain and interface pLDDT, and mean interface PAE; Supplementary Table S4).

As shown in **Figure 2A**, the Axl–Gas6 complexes from all three species adopt a highly conserved global architecture. The relative arrangement of the receptor Ig-like domains and the ligand LG domains is preserved across human, mouse, and zebrafish, indicating that the overall mode of TAM–ligand engagement is evolutionarily conserved despite substantial sequence divergence.

To further evaluate interspecies differences at a finer structural level, we performed direct structural superpositions of the RBD complexes for all biologically relevant TAM–ligand pairs (Axl–Gas6, Mertk–Gas6, and Tyro3–Protein S). As shown in **Figure 2B**, receptor backbones overlap closely across species in all complexes, confirming strong conservation of the receptor scaffold. In contrast, subtle but reproducible differences in ligand positioning and orientation are observed between mammalian and zebrafish complexes. These differences are most pronounced in the Tyro3–Protein S complex, where the zebrafish ligand exhibits a distinct orientation relative to the conserved receptor framework.

Together, these results demonstrate that TAM receptor–ligand complexes maintain a conserved overall architecture across species, while exhibiting modest species-specific variations in ligand orientation. These structural differences, though subtle, suggest potential modulation of interface geometry and interaction chemistry in regenerative versus non-regenerative contexts, motivating subsequent quantitative analyses of binding interfaces and energetics.

**Figure 2. Conserved global architecture and subtle species-specific variation in TAM receptor–ligand RBD complexes.** **(A)** AlphaFold-Multimer models of the Axl–Gas6 receptor-binding domain (RBD) complexes from *Homo sapiens*, *Mus musculus*, and Danio rerio. Only the extracellular ligand-binding regions are shown, comprising Ig-like domains 1–2 of the receptor (blue) and LG domains 1–2 of the ligand Gas6 (red). All complexes adopt a highly similar overall architecture across species. **(B)** Structural superposition of TAM–ligand RBD complexes across species for Axl–Gas6, Mertk–Gas6, and Tyro3–Protein S. Human, mouse, and zebrafish complexes are shown in distinct colors. While the receptor scaffolds overlap closely, subtle species-specific differences in ligand orientation are observed, most prominently in the Tyro3–Protein S complex, as indicated by the arrow.

### 3.3 Interface architecture and binding energetics of TAM–ligand complexes across species

To determine whether the observed species-specific differences in ligand positioning translate into measurable changes at the interaction interface, we next performed a quantitative analysis of receptor–ligand binding interfaces for all modeled TAM–ligand receptor-binding domain (RBD) complexes. Interface properties were evaluated using PDBePISA, focusing on buried surface area, predicted interface binding free energy (ΔiG), and the network of stabilizing non-covalent interactions.

Across all complexes, TAM–ligand interfaces exhibited substantial buried surface areas consistent with stable protein–protein interactions. For Axl–Gas6, interface areas were broadly comparable across species, with mammalian complexes displaying slightly larger buried surfaces than the zebrafish complex. In contrast, more pronounced interspecies variation was observed for the Mertk–Gas6 and Tyro3–Protein S interfaces, indicating greater plasticity in these interaction surfaces.

Analysis of predicted interface binding free energies revealed a general tendency toward more favorable (more negative) energetics in zebrafish complexes relative to their mammalian counterparts, although this trend was not uniform across all receptor–ligand pairs. The energetic differences were modest for Axl–Gas6, intermediate for Mertk–Gas6, and most pronounced for Tyro3–Protein S, suggesting that the species-specific ligand reorientation observed in Figure 2 is differentially coupled to interface stabilization across TAM receptors. Notably, the predicted interface free energy ($\Delta iG$) for the Axl–Gas6 complex in Danio rerio was slightly positive, reflecting limitations of static, interface-based energy estimation applied to isolated receptor-binding domains rather than the absence of biologically relevant interactions.

Interfacial contact analysis further refined these trends. While the total number of hydrogen bonds was broadly conserved across species for all complexes, zebrafish assemblies exhibited an increased number of salt bridges and charged interactions at the interface. This enrichment was uneven across receptor–ligand pairs and was pronounced in the Tyro3–Protein S interaction, consistent with its larger structural divergence relative to mammalian counterparts.

When normalized by interface area (Figure **3E–F**), salt-bridge density—but not hydrogen-bond density—shows a selective enrichment in zebrafish complexes, indicating tighter electrostatic packing rather than a global increase in interfacial contacts. In contrast, hydrogen-bond density remains broadly comparable across species (Figure **3F**), indicating that differences in interface organization are not driven by a general increase in polar contacts but are instead selectively associated with electrostatic (salt-bridge) enrichment.

In the Axl–Gas6 interaction, salt-bridge density is elevated in zebrafish relative to human complexes (Figure 3E), indicating increased electrostatic interaction density at the interface, without implying a direct relationship to regenerative capacity. However, mouse Axl–Gas6 complexes display similarly high salt-bridge density, despite lacking regenerative capacity. This suggests that increased electrostatic density in Axl–Gas6 reflects a general enhancement of interface sensitivity rather than a regeneration-specific optimization, and occurs without major ligand reorientation or localized electrostatic hotspot clustering.

Differences in predicted binding energetics did not arise from a single dominant interaction but instead emerged from the cumulative effect of small-scale changes in interface geometry and residue composition. These findings indicate that subtle shifts in ligand orientation, rather than major rearrangements of the conserved receptor scaffold, are sufficient to modulate TAM–ligand interface properties across species.

Together, these results demonstrate that conserved TAM receptor architectures support species-specific tuning of ligand-binding interfaces. In regenerative species such as Danio rerio, this tuning is characterized by enhanced electrostatic contributions and, in selected receptor–ligand pairs, more favorable predicted interface energetics, providing a comparative structural framework to interpret differences in TAM receptor–ligand interface organization across vertebrates.

**Figure 3. Quantitative comparison of binding interface properties and predicted energetics of TAM–ligand complexes across species.** (**A**) Buried interface area for Axl–Gas6, Mertk–Gas6, and Tyro3–Protein S receptor-binding domain (RBD) complexes from Homo sapiens, Mus musculus, and Danio rerio. (**B**) Predicted interface binding free energy ($\Delta iG$) for each complex. (**C**) Number of hydrogen bonds (NHB) at the receptor–ligand interface. (**D**) Number of salt bridges (NSB) at the interface.

(**E**) Salt-bridge density ($\rho\_SB$), defined as the number of salt bridges normalized by buried interface area (per 1000 $Å^2$).

(**F**) Hydrogen-bond density ($\rho\_HB$), defined as the number of hydrogen bonds normalized by buried interface area (per 1000 $Å^2$).

Absolute contact counts (**3C–D**) reflect overall interaction numbers, whereas normalized densities (**3E–F**) capture differences in interaction packing and electrostatic organization independent of interface size. Values correspond to single representative receptor–ligand complexes and are interpreted comparatively rather than statistically.
Values shown correspond to single, representative receptor–ligand complexes and do not reflect population-level sampling; therefore, differences are interpreted comparatively and mechanistically rather than through statistical inference. Together, these analyses show that while interface area and hydrogen bonding are broadly conserved across species, differences in predicted binding energetics are associated with increased electrostatic interactions in Danio rerio TAM–ligand complexes.

In contrast, the Tyro3–Protein S complex uniquely combines elevated salt-bridge density with pronounced ligand reorientation and localized electrostatic clustering, distinguishing it from Axl–Gas6 and Mertk–Gas6 interfaces and motivating focused residue-level analysis in the following section.

### 3.4. Residue-level interaction patterns at TAM receptor–ligand interfaces

To characterize the molecular determinants underlying TAM receptor–ligand recognition, we analyzed residue-level hydrogen bonds and salt bridges at the interfaces of the Axl–Gas6, Mertk–Gas6, and Tyro3–Protein S complexes across Homo sapiens, Mus musculus, and Danio rerio orthologs.

Across all three receptor–ligand pairs, the interfaces are stabilized by a combination of polar hydrogen bonds and electrostatic salt bridges, consistent with the energetic trends reported above. While the total number of interfacial contacts is broadly comparable across species, the identity, geometry, and distribution of individual interactions differ substantially, reflecting species-specific remodeling of local contact networks rather than changes in global interface architecture.

In the Axl–Gas6 complexes, several conserved polar interactions are observed across species, particularly involving charged residues within the receptor Ig-like domains and the ligand LG domains. Notably, the Danio rerio complex exhibits an increased contribution of electrostatic contacts, including multiple salt bridges formed by acidic residues on Gas6 engaging basic residues on Axl. Several of these interactions involve repeated contacts between the same residue pairs through distinct atomic geometries, suggesting increased interaction redundancy and local stabilization of the interface. This pattern is consistent with the increased electrostatic contribution and interaction redundancy observed for the Danio rerio Axl–Gas6 interface.

For the Mertk–Gas6 interface, hydrogen bonds dominate the interaction network in Homo sapiens and Mus musculus, primarily involving backbone–sidechain and polar sidechain interactions. In contrast, the Danio rerio complex displays a higher relative contribution of salt bridges, largely mediated by conserved acidic residues on Gas6 engaging basic residues on Mertk. These findings suggest that subtle shifts in charge complementarity, rather than large-scale rearrangements of the binding mode, contribute to species-specific stabilization of the Mertk–Gas6 interaction.

The Tyro3–Protein S complexes display the greatest divergence in residue-level interaction patterns across species. Although the overall binding mode remains conserved, the specific hydrogen bonds and salt bridges differ markedly, particularly in Danio rerio, where additional electrostatic interactions form distinct charged clusters at the interface. These differences align with the ligand reorientation observed in structural superpositions and may reflect reduced evolutionary constraint on this receptor–ligand pair, allowing greater structural and electrostatic plasticity compared to Axl–Gas6 and Mertk–Gas6.

Together, these analyses indicate that species-specific differences in TAM receptor–ligand binding are encoded not in global interface architecture but in fine-scale remodeling of residue-level contact networks. In regenerative species such as Danio rerio, enrichment of electrostatic interactions and interaction redundancy appears to enhance interface stabilization, providing a potential structural basis for more efficient TAM signaling in regenerative neuroimmune contexts. A complete listing of all hydrogen bonds and salt bridges, including residue identities and interaction distances, is provided in Supplementary Tables S1–S3.

### 3.5 Electrostatic hotspots at the Tyro3–Protein S interface (Homo sapiens vs Danio rerio)

Electrostatic interaction hotspots at the Tyro3–Protein S interface were identified comparatively between Homo sapiens and Danio rerio by integrating APBS-derived electrostatic potential differences with interface contact analysis, and were subsequently mapped onto the three-dimensional structures of both complexes (Figure 4). Hotspots were defined as receptor–ligand residue pairs with high mean absolute electrostatic potential differences ($|\Delta\phi|$) and at least one recurrent interfacial contact.

Structural mapping showed that, in both species, electrostatic hotspots are not uniformly distributed across the protein surfaces but instead concentrate within a defined binding region. On the receptor side, hotspot residues cluster predominantly within the extracellular Ig1–Ig2 domains of Tyro3, forming a contiguous electrostatic interaction patch at the receptor–ligand interface in both Homo sapiens and Danio rerio (Figure 4A,B). These sites comprise acidic, basic, and polar residues that collectively shape the electrostatic landscape of the binding surface.

On the ligand side, hotspot residues localize mainly within the laminin G–like (LG) domain of Protein S, again in both species, consistent with its role in receptor recognition. The arrangement of charged and polar residues in the LG region supports electrostatic complementarity with the Ig1–Ig2 surface of Tyro3, indicating a charge-driven contribution to interface stabilization across the two orthologous complexes.

Quantitatively, residue pairs exhibiting the largest $|\Delta\phi|$ values frequently coincide with multiple interfacial contacts (Figure 4C), suggesting that stronger electrostatic mismatches tend to occur in recurrent contact contexts rather than as isolated interactions. Moreover, several hotspot residues participate in multiple top-ranking pairs (Figure 4D), consistent with hotspot "clusters" that concentrate electrostatic contributions into a limited interfacial area.

Overall, while the specific identities of top-ranked residue pairs differ between Homo sapiens and Danio rerio, the spatial organization of electrostatic hotspots is conserved:

both complexes display a shared electrostatic interaction core centered on the Tyro3 Ig1–Ig2 domains and the Protein S LG region (Figure 4A,B,D). This cross-species convergence supports electrostatic complementarity as a conserved structural determinant of Tyro3–Protein S binding.

**Figure 4. Structural localization and quantitative characterization of electrostatic hotspots at the Tyro3–Protein S interface**

**(A)** Surface representation of the Tyro3–Protein S complex in Homo sapiens, highlighting electrostatic interaction hotspots at the receptor–ligand interface. Hotspot residues are predominantly localized within the Ig1–Ig2 domains of Tyro3 (receptor, light blue) and the laminin G–like (LG) domain of Protein S (ligand, dark blue). Electrostatic hotspot residues on the receptor and ligand are shown in yellow and orange, respectively. The inset shows a zoomed view of the interface region, illustrating the spatial clustering of high-impact electrostatic interactions.

**(B)** Equivalent structural mapping of electrostatic hotspots in Danio rerio, revealing a similar spatial organization of hotspot residues at the Ig1–Ig2/LG interface despite species-specific differences in residue identity. The zoomed inset highlights the conserved interfacial hotspot region.

**(C)** Relationship between the mean absolute electrostatic potential difference ($|\Delta\phi|$) and interface contact multiplicity for hotspot residue pairs in Homo sapiens (blue) and Danio rerio (orange). Residue pairs with high $|\Delta\phi|$ values frequently engage in multiple interfacial contacts, indicating that dominant electrostatic interactions are reinforced by structural recurrence.

**(D)** Ranking of the top electrostatic interface residue pairs based on mean $|\Delta\phi|$ values using full-length residue numbering. While individual residue identities differ between species, the distribution of high-ranking interactions supports the presence of a conserved electrostatic interaction core at the Tyro3 Ig1–Ig2 and Protein S LG interface.

**3.6 Conservation of electrostatic interface hotspots between Homo sapiens and Danio rerio**

To evaluate whether the strongest electrostatic hotspots identified at the Tyro3–Protein S interface are evolutionarily conserved, we performed a comparative sequence-based analysis between Homo sapiens and Danio rerio (Figure 5). Full-length amino acid sequences of Tyro3 (receptor) and Protein S (ligand) were aligned independently, and interface residues previously identified from PDBePISA and APBS analyses were mapped onto full-length sequence numbering.

Electrostatic hotspot residues were defined as those contributing to the top ten receptor–ligand residue pairs ranked by the mean absolute electrostatic potential difference ($|\Delta\phi|$). Residues participating in multiple high-ranking interactions were treated as components of electrostatic interaction clusters rather than isolated contacts.

Sequence alignment revealed that a substantial fraction of hotspot residues in Tyro3 are conserved or functionally conserved between Homo sapiens and Danio rerio (Figure 5A). In particular, residues located within the Ig1–Ig2 domains frequently align to positions retaining similar physicochemical properties, most notably acidic or basic side chains, indicating preservation of electrostatic function even when strict residue identity is not maintained.

A comparable pattern was observed for Protein S (Figure 5B), where hotspot residues localized to the laminin G–like (LG) domain showed conservation at the level of charge and polarity across species. Several ligand-side hotspots correspond to aligned positions that maintain electrostatic complementarity with the receptor interface, supporting a conserved role of the LG domain in Tyro3 recognition.

Importantly, although individual receptor–ligand residue pairs are not always conserved one-to-one between species, the global electrostatic architecture of the interface is preserved. Both complexes display aligned charged patches on Tyro3 and Protein S that correspond to regions of high electrostatic mismatch identified by APBS. This indicates evolutionary rewiring of specific contacts while maintaining an overall conserved electrostatic interaction framework.

Together, these results demonstrate that electrostatic hotspots at the Tyro3–Protein S interface are conserved at the level of structural location and electrostatic function in both Homo sapiens and Danio rerio. This conservation supports the existence of an evolutionarily preserved, charge-driven mechanism underlying Tyro3–Protein S binding.

**Figure 5. Conservation of electrostatic interface hotspots between Homo sapiens and Danio rerio.**

Multiple sequence alignment of full-length Tyro3 **(A)** and Protein S **(B)** sequences from Homo sapiens and Danio rerio. Electrostatic hotspot residues identified from APBS and PDBePISA analyses and belonging to the top-ranked receptor–ligand residue pairs (by mean $|\Delta\phi|$) are highlighted on the alignments. Conservation scores and consensus sequences are shown below each alignment. Hotspot residues display partial to high conservation across species, particularly within the Tyro3 Ig1–Ig2 domains and the laminin G–like (LG) domain of Protein S. While strict residue identity is not always preserved, the majority of aligned hotspot positions maintain similar physicochemical properties, indicating functional conservation of electrostatic interactions underlying Tyro3–Protein S recognition.

## 4 Discussion

In this study, we provide a comparative structural and biophysical analysis of TAM receptor–ligand interactions across regenerative (Danio rerio) and non-regenerative (Homo sapiens and Mus musculus) vertebrates. Our results reveal that evolutionary divergence in TAM signaling efficiency is not driven by changes in overall receptor–ligand architecture, but rather by fine-scale remodeling of interface electrostatics and residue-level interaction networks within a conserved structural scaffold.

### 4.1 Conserved architecture with divergent molecular tuning

Sequence analyses confirm substantial divergence between zebrafish and mammalian TAM receptors and ligands, particularly within ligand-binding regions. However, structural modeling demonstrates that the Ig-like domains of TAM receptors and the LG domains of Gas6 and Protein S retain strong fold conservation across species. This decoupling between sequence divergence and structural conservation indicates strong evolutionary constraint on the global mode of TAM–ligand engagement.

Importantly, while the receptor scaffolds overlap closely across species, subtle but consistent differences in ligand orientation are observed, most prominently in the Tyro3–Protein S complex. These differences do not disrupt binding but instead reshape the local interface geometry, suggesting that evolutionary adaptation operates through modulation of interaction topology rather than through wholesale changes in binding mode.

Notably, this pattern coexists with pronounced divergence in the zebrafish Protein S ligand-binding region, which shows substantially reduced structural similarity to mammalian counterparts. The apparent preservation of receptor engagement despite this divergence raises the question of which molecular features—beyond global fold conservation—stabilize TAM–ligand interfaces across species.

### 4.2 Electrostatic enrichment as a hallmark of regenerative TAM interfaces

Quantitative interface analyses reveal that zebrafish TAM–ligand complexes tend to exhibit enhanced electrostatic contributions, reflected in increased salt-bridge counts and more favorable predicted interface energetics in selected receptor–ligand pairs. To disentangle whether this enrichment reflects a genuine increase in electrostatic packing rather than differences in interface size, salt-bridge and hydrogen-bond counts were normalized by buried interface area. Salt-bridge density ($\rho\_SB$) and hydrogen-bond density ($\rho\_HB$) were defined as the number of interactions per 1000 $Å^2$ of interface surface (see 2.7.1). This normalization enables direct cross-species comparison of interaction organization independent of interface extent. This trend is especially pronounced for Tyro3–Protein S, which displays both the largest structural divergence and the strongest electrostatic enrichment among the complexes analyzed.

Residue-level analysis shows that these energetic differences arise from the cumulative effect of multiple small-scale changes rather than from a single dominant interaction. Zebrafish interfaces frequently display interaction redundancy, where the same residues engage in multiple hydrogen bonds or salt bridges through alternative atomic geometries. Such redundancy may increase the robustness of ligand binding under fluctuating physiological conditions, a feature potentially advantageous in regenerative environments characterized by dynamic inflammation and tissue remodeling. In particular, redundant electrostatic contacts could buffer transient perturbations in ionic strength, pH, or local conformational variability, helping stabilize receptor engagement when individual interactions are intermittently weakened.

Importantly, the observation of a slightly positive predicted interface free energy for the Danio rerio Axl–Gas6 complex does not imply weak or non-functional binding. PDBePISA-derived ΔiG values capture only static, geometry-based contributions from truncated receptor-binding domains and do not account for membrane anchoring, receptor clustering, multivalency, or dynamic electrostatic stabilization. In interfaces dominated by electrostatic interactions, binding robustness may therefore emerge from interaction redundancy and spatial charge organization rather than from strongly favorable static binding energies.

Together, these findings suggest that enhanced electrostatic contributions and interaction redundancy provide a plausible mechanistic interpretation by which TAM–ligand interfaces may accommodate substantial ligand divergence while preserving receptor engagement within a conserved structural scaffold.

### 4.3 Conserved electrostatic hotspot architecture despite evolutionary rewiring

APBS-based electrostatic mapping identifies clustered interface hotspots at the Tyro3–Protein S interface that are conserved in spatial organization across species. Although the exact residue identities differ between human and zebrafish, the majority of hotspot positions preserve their physicochemical roles, maintaining charge complementarity and electrostatic mismatch at corresponding interface regions.

This pattern supports a model in which TAM interfaces evolve through functional conservation of electrostatic landscapes rather than strict residue conservation. Evolutionary rewiring of individual contacts preserves a shared electrostatic interaction core, allowing species-specific optimization while maintaining reliable receptor–ligand recognition. More broadly, this conservation of electrostatic architecture despite residue-level turnover is consistent with evolutionary constraint acting on interface function and charge complementarity rather than on specific amino acid identities.

### 4.4 Implications for regenerative neuroimmune signaling

TAM receptors play a central role in phagocytosis, efferocytosis, and immune resolution in the CNS. Enhanced electrostatic complementarity and interaction redundancy at zebrafish TAM interfaces may lower activation thresholds or increase signaling persistence, which may bias TAM signaling toward more persistent or robust engagement in regenerative contexts, consistent with enhanced debris clearance observed in zebrafish. In such contexts, interaction redundancy may buffer transient environmental and molecular fluctuations during inflammation and tissue remodeling, stabilizing signaling output and supporting sustained phagocytic engagement.

From a translational perspective, our results suggest that engineering TAM receptor–ligand interfaces to enhance electrostatic complementarity and stabilize key interaction clusters—rather than altering global binding modes—may represent a viable strategy to explore the modulation of repair-oriented immune responses in the mammalian nervous system.

### 5 Limitations

Several limitations should be considered when interpreting the results of this study. First, all receptor–ligand complexes were modeled computationally and analyzed as static structures. While AlphaFold-Multimer provides high-confidence predictions, dynamic effects such as conformational flexibility, induced fit, and long-timescale rearrangements are not explicitly captured.

Second, predicted interface binding energies derived from PDBePISA represent approximate estimates based on isolated receptor-binding domains rather than full-length, membrane-embedded receptors. Consequently, absolute $\Delta iG$ values should be interpreted comparatively rather than quantitatively.

Third, molecular dynamics simulations were not performed, limiting our ability to assess dynamic stability, solvent effects, and entropy-driven contributions to binding. Additionally, downstream signaling events, receptor clustering, and membrane context were not modeled and may further modulate TAM activation in vivo.

Fourth, a limitation of this study is that the regenerative comparison is restricted to a single regenerative vertebrate species, Danio rerio, because comparable TAM receptor–ligand sequence information required to define interaction domains and model the complexes was not available for additional regenerative taxa within the scope of this work.

Finally, although zebrafish serves as a powerful model of CNS regeneration, species-specific differences beyond TAM signaling, including immune cell composition and transcriptional programs, also contribute to regenerative capacity and are not addressed here.

## 6 Proposed Experimental Validation and Future Directions

The structural and electrostatic features identified in this study generate several experimentally testable hypotheses:

- Targeted mutagenesis

Introduce zebrafish-like charged residues or electrostatic clusters into mammalian TAM receptors or ligands to assess their impact on ligand affinity and receptor activation.

- Biophysical binding assays

Measure binding affinities of wild-type and mutant TAM–ligand pairs using surface plasmon resonance (SPR) or biolayer interferometry (BLI) to validate predicted electrostatic contributions.

- Cellular phagocytosis assays

Evaluate whether engineered TAM variants enhance microglial or astrocytic phagocytosis of apoptotic material in vitro.

- In vivo functional assays

Test whether electrostatically optimized TAM signaling improves debris clearance and functional recovery in mammalian models of CNS injury.

- Molecular dynamics simulations

Extend the present work with atomistic simulations to quantify dynamic stability and electrostatic persistence at TAM interfaces.

## 7 Conclusions

Taken together, our results support a coherent model in which TAM receptor–ligand interactions are governed by a conserved structural scaffold that tolerates extensive sequence divergence through electrostatic and geometric compensation. In regenerative species such as Danio rerio, this compensation manifests not as uniformly stronger static binding, but as enhanced electrostatic organization and interaction redundancy at the interface, which may stabilize receptor activation under fluctuating physiological conditions and are consistent with conditions that may support more robust phagocytic signaling and immune resolution in regenerative species.

By revealing how evolutionary rewiring preserves electrostatic function while allowing species-specific optimization, our work establishes a structural bridge between molecular interaction chemistry and regenerative neuroimmune outcomes. These insights lay a rational foundation for engineering TAM signaling pathways to promote repair-oriented immune responses in the mammalian CNS.

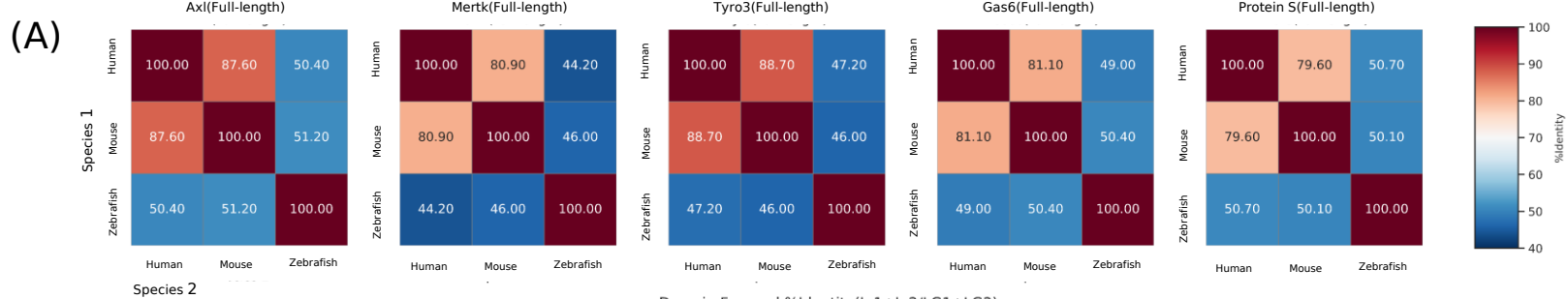

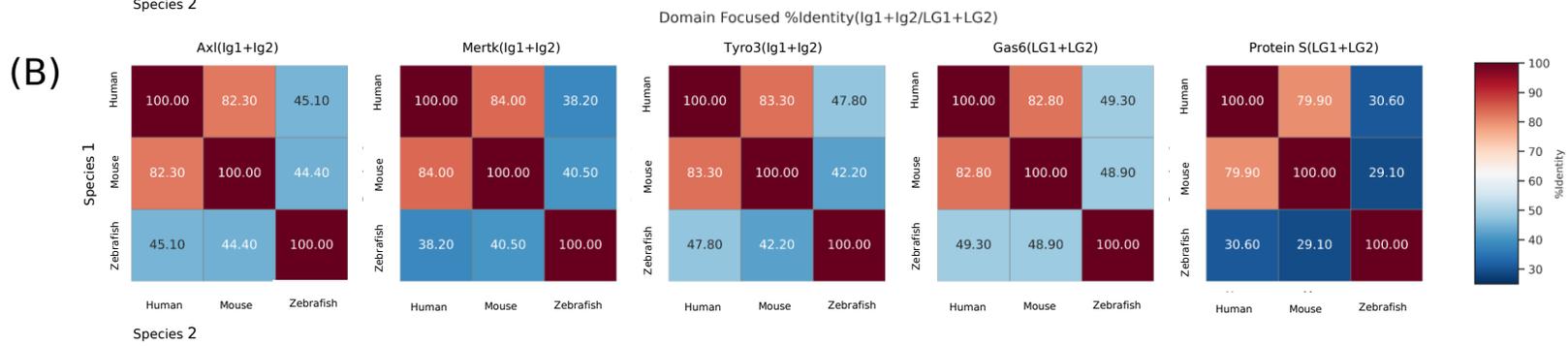

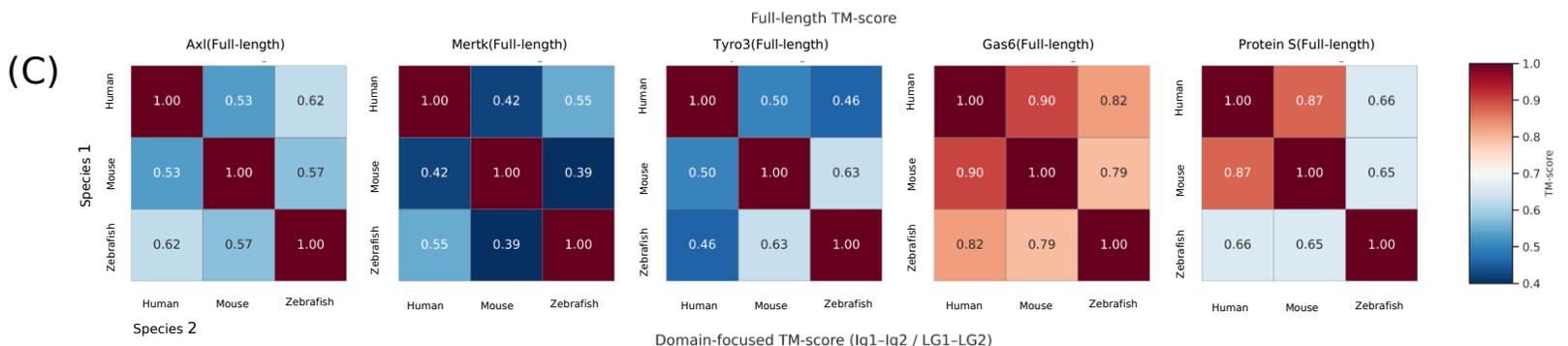

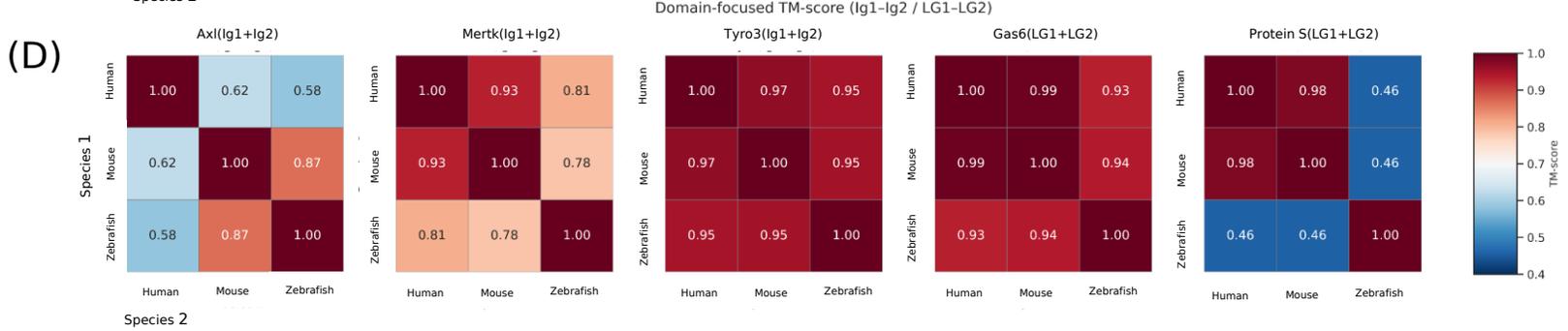

**(A)**

Axl_Gas6_Human

Axl_Gas6_Mouse

Axl_Gas6_Zebrafish

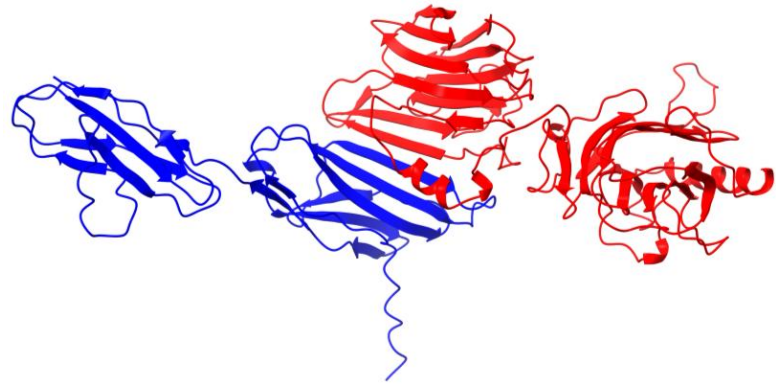
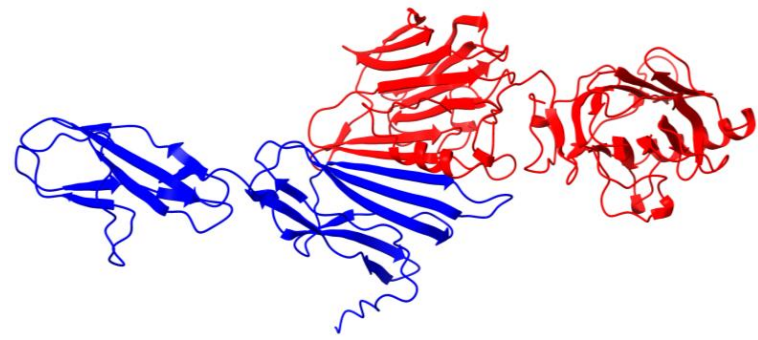
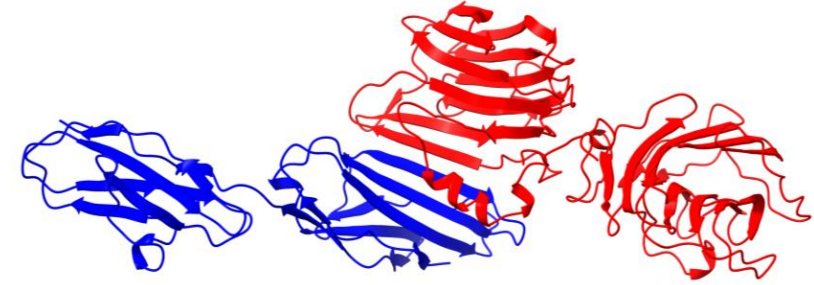

Receptor Axl
(Ig1+Ig2)

Ligand Gas6
(LG1+LG2)

**(B)**

Structural superposition of the Axl – Gas6
complex across species

Structural superposition of the Mertk– Gas6
complex across species

Structural superposition of the Tyro3–
Protein S complex across species

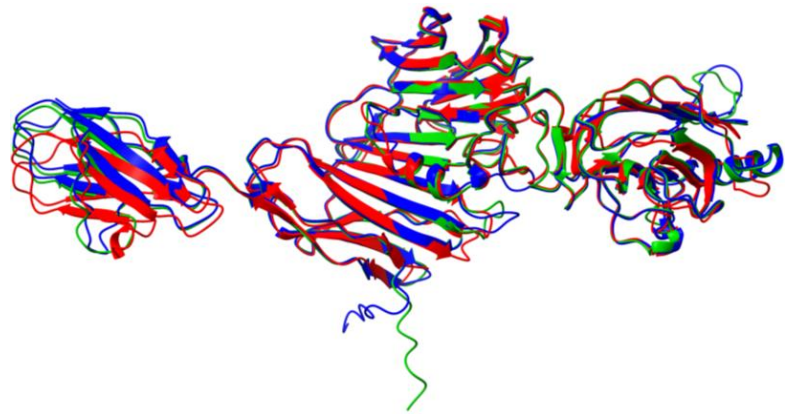
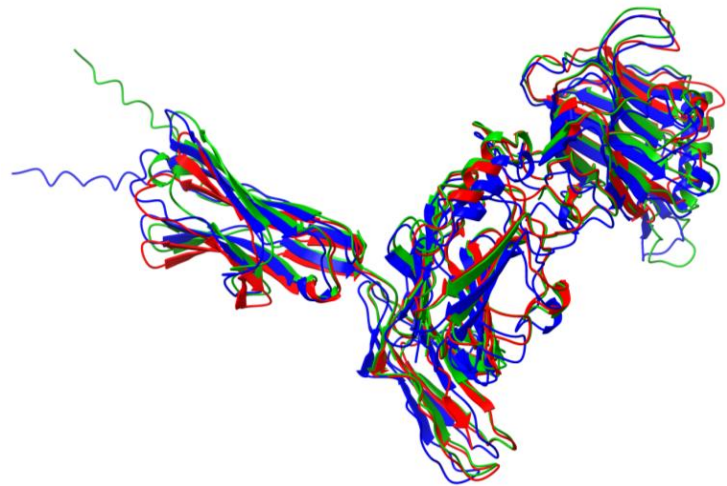
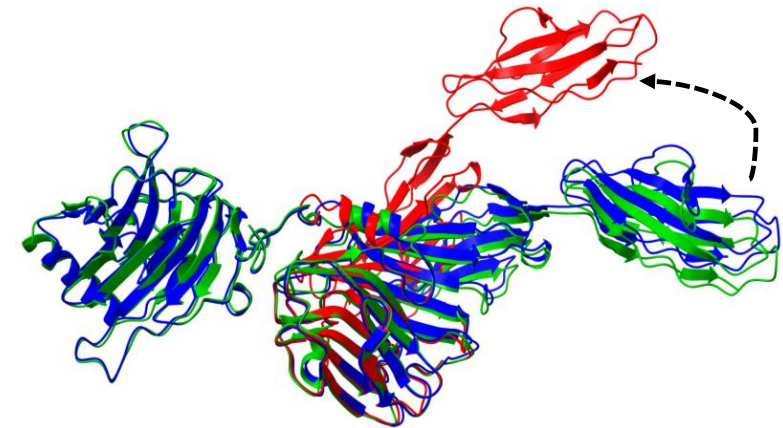

Human    Mouse    Zebrafish

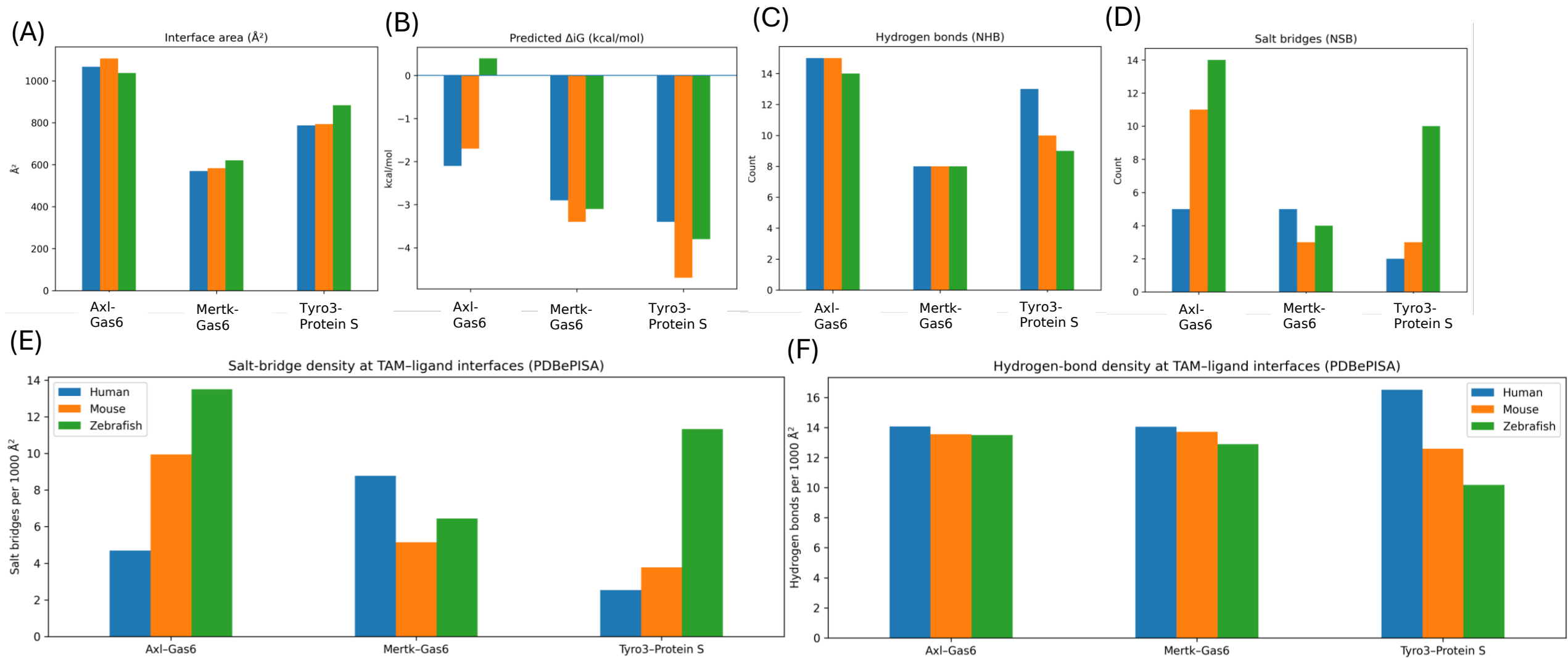

Figure 3

**(A)** Electrostatic hotspots at the Tyro3 (Ig1–Ig2) domains and the Protein S (LG) region (Human)

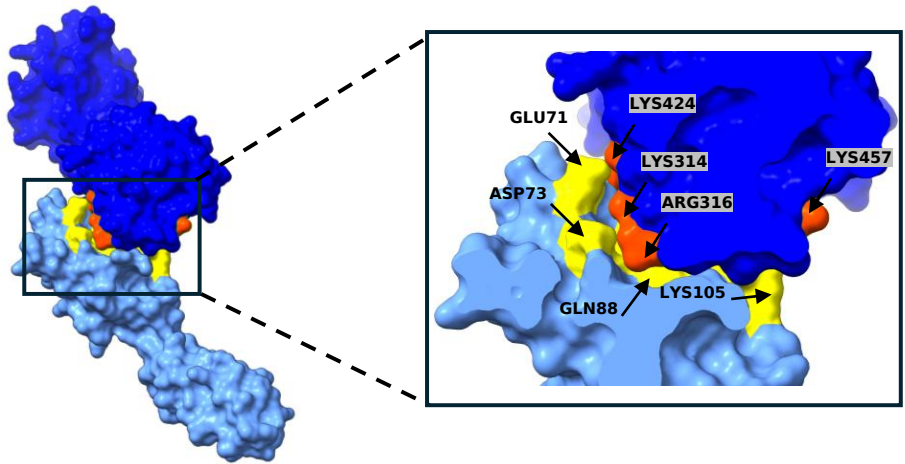

Ligand    Receptor    Hotspot Ligand    Hotspot Receptor

**(B)** Electrostatic hotspots at the Tyro3 (Ig1–Ig2) domains and the Protein S (LG) region (Zebrafish)

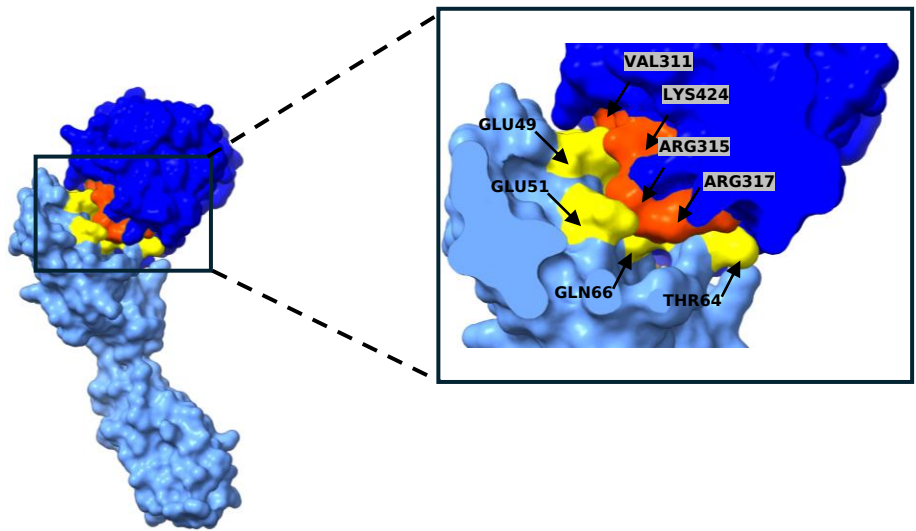

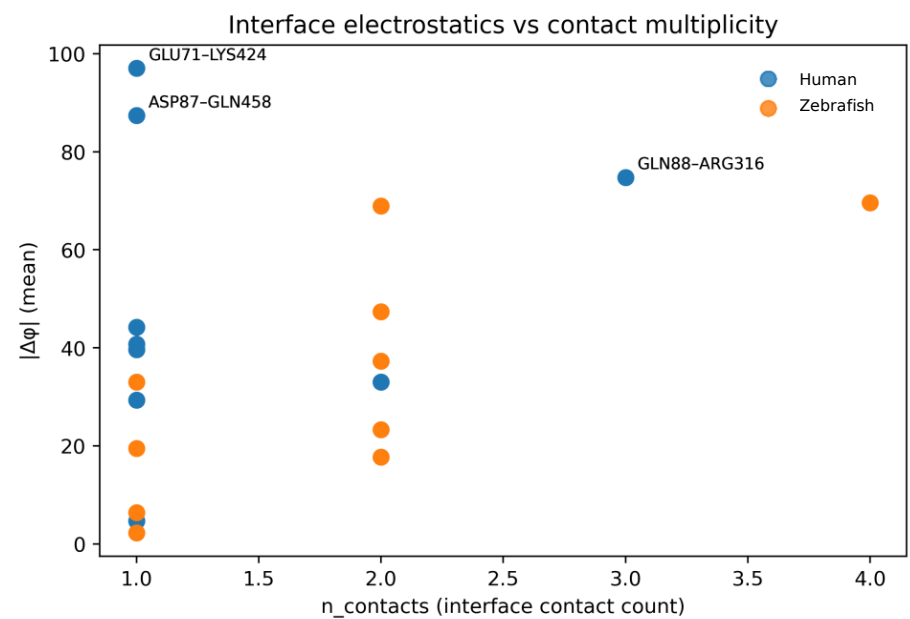

**(D)**

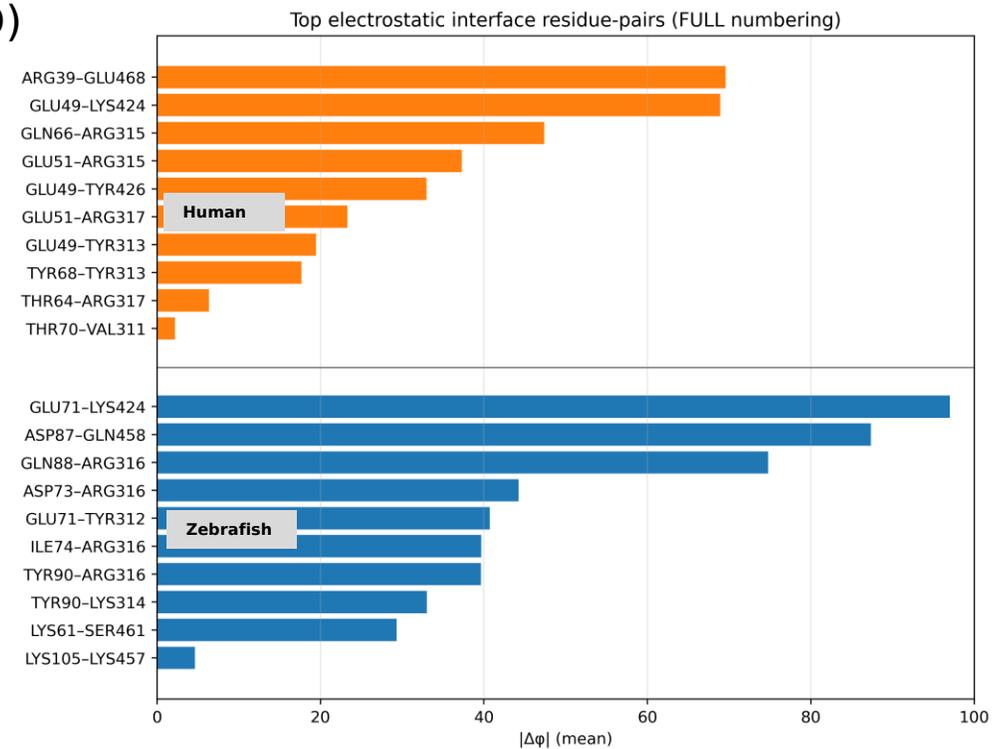

## (A) Sequence conservation of electrostatic interface hotspots in Tyro3 (Human vs Zebrafish)

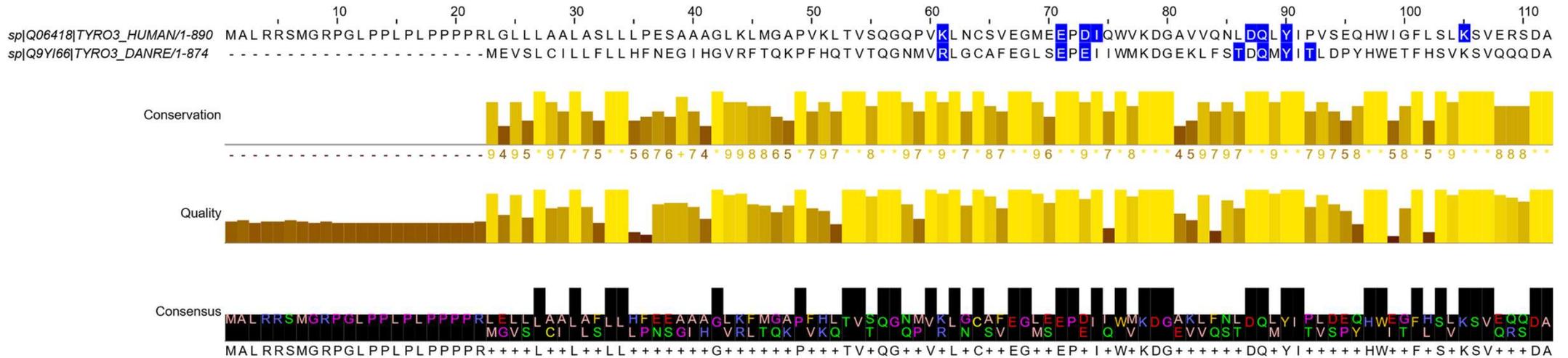

## (B) Sequence conservation of electrostatic interface hotspots in Protein S (Human vs Zebrafish)

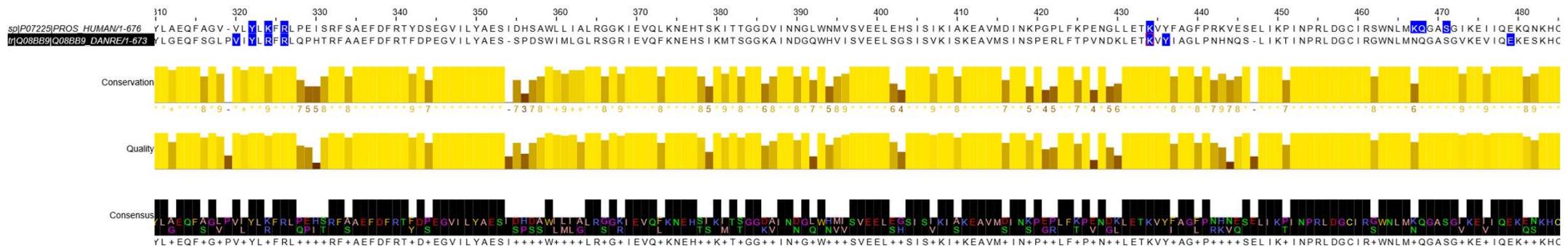

Electrostatic interface hotspots identified by APBS-based electrostatic analysis.

# Supplementary Table S1

Table 1: **Table S1.** Hydrogen bonds and salt bridges at the Axl–Gas6 interface in Homo sapiens, Mus musculus, and Danio rerio complexes. Distances are reported in Å.

| Species | Axl residue | Gas6 residue | Distance (Å) |
|---|---|---|---|
| **Hydrogen bonds** | | | |
| Homo | ARG 55 (NH1) | GLU 460 (OE2) | 2.90118 |
| Homo | ARG 55 (NH2) | THR 457 (OG1) | 3.47037 |
| Homo | ARG 55 (NH2) | GLU 460 (OE2) | 2.87533 |
| Homo | SER 81 (N) | LYS 312 (O) | 2.97996 |
| Homo | GLN 83 (N) | ARG 310 (O) | 2.85917 |
| Homo | THR 84 (OG1) | ASP 455 (OD2) | 2.53754 |
| Homo | GLN 85 (N) | ARG 308 (O) | 2.83595 |
| Homo | GLU 63 (OE1) | ARG 308 (NH1) | 3.20772 |
| Homo | GLU 66 (OE1) | LYS 312 (NZ) | 2.94200 |
| Homo | ASP 80 (OD1) | LEU 314 (N) | 2.64616 |
| Homo | SER 81 (O) | LYS 312 (N) | 2.87474 |
| Homo | SER 81 (O) | ARG 313 (NH1) | 2.95783 |
| Homo | THR 82 (OG1) | ARG 313 (NH2) | 2.91883 |
| Homo | GLN 83 (O) | ARG 310 (N) | 2.83942 |
| Homo | GLN 85 (O) | ARG 308 (N) | 2.98673 |
| Mus | ARG 49 (NH1) | GLU 457 (OE2) | 2.60834 |
| Mus | ASN 75 (N) | LYS 309 (O) | 2.94988 |
| Mus | THR 76 (OG1) | GLU 451 (OE2) | 2.00402 |
| Mus | GLN 77 (N) | ARG 307 (O) | 2.84110 |
| Mus | GLN 79 (N) | ARG 305 (O) | 2.75491 |
| Mus | GLN 95 (NE2) | SER 453 (OG) | 3.12766 |
| Mus | GLU 57 (OE1) | ARG 305 (NH1) | 2.70586 |
| Mus | GLU 60 (OE1) | LYS 309 (NZ) | 3.16060 |
| Mus | ASP 74 (OD1) | LEU 311 (N) | 2.60433 |
| Mus | ASN 75 (O) | LYS 309 (N) | 2.79181 |
| Mus | ASN 75 (O) | ARG 310 (NH1) | 3.23600 |
| Mus | THR 76 (OG1) | ARG 310 (NH2) | 2.85567 |
| Mus | GLN 77 (O) | ARG 307 (N) | 2.77086 |
| Mus | GLN 79 (O) | ARG 305 (N) | 2.92177 |
| Mus | PRO 81 (O) | ARG 296 (NH1) | 3.33290 |
| Danio | THR 99 (N) | ARG 301 (O) | 3.06682 |
| Danio | GLN 101 (N) | ARG 299 (O) | 2.83132 |
| Danio | GLN 103 (N) | ARG 297 (O) | 2.77924 |
| Danio | GLU 95 (O) | ARG 301 (NH2) | 2.87480 |
| Danio | ASP 98 (OD1) | ARG 303 (N) | 2.86406 |
| Danio | ASP 98 (OD1) | ARG 303 (NE) | 3.83540 |
| Danio | THR 99 (O) | ARG 301 (N) | 2.86939 |
| Danio | THR 99 (O) | ARG 302 (NH1) | 3.17789 |

| Species | Axl residue | Gas6 residue | Distance ( Å ) |
|---------|-------------|--------------|----------------|
| Danio | ASN 100 (OD1) | ARG 302 (NH1) | 2.69914 |
| Danio | ASN 100 (OD1) | ARG 302 (NH2) | 2.66501 |
| Danio | GLN 101 (O) | ARG 299 (N) | 2.84676 |
| Danio | GLN 101 (OE1) | ARG 299 (NH2) | 3.47620 |
| Danio | GLN 103 (O) | ARG 297 (N) | 2.88018 |
| Danio | LEU 105 (O) | ARG 288 (NH1) | 2.84905 |

**Salt bridges**

| Species | Axl residue | Gas6 residue | Distance ( Å ) |
|---------|-------------|--------------|----------------|
| Homo | ARG 55 (NH1) | GLU 460 (OE1) | 3.43849 |
| Homo | ARG 55 (NH1) | GLU 460 (OE2) | 2.90118 |
| Homo | ARG 55 (NH2) | GLU 460 (OE2) | 2.87533 |
| Homo | GLU 63 (OE1) | ARG 308 (NH1) | 3.20772 |
| Homo | GLU 66 (OE1) | LYS 312 (NZ) | 2.94200 |
| Mus | ARG 49 (NH1) | GLU 457 (OE1) | 3.15519 |
| Mus | ARG 49 (NH1) | GLU 457 (OE2) | 2.60834 |
| Mus | ARG 49 (NH2) | GLU 457 (OE2) | 2.86811 |
| Mus | GLU 57 (OE1) | ARG 305 (NH1) | 2.70586 |
| Mus | GLU 57 (OE2) | ARG 305 (NH1) | 3.91468 |
| Mus | GLU 60 (OE1) | LYS 309 (NZ) | 3.16060 |
| Mus | GLU 60 (OE1) | ARG 307 (NH2) | 2.91502 |
| Mus | GLU 60 (OE1) | ARG 307 (NH1) | 2.94510 |
| Mus | GLU 60 (OE2) | ARG 307 (NH1) | 3.38560 |
| Mus | GLU 84 (OE2) | LYS 464 (NZ) | 2.42078 |
| Mus | ASP 85 (OD2) | HIS 658 (NE2) | 3.79305 |
| Danio | GLU 81 (OE1) | ARG 297 (NE) | 2.00100 |
| Danio | GLU 81 (OE2) | ARG 297 (NE) | 3.55728 |
| Danio | GLU 81 (OE2) | ARG 297 (NH2) | 3.77935 |
| Danio | ASP 84 (OD1) | ARG 299 (NH1) | 2.99200 |
| Danio | ASP 84 (OD1) | ARG 299 (NH2) | 3.42945 |
| Danio | ASP 84 (OD1) | ARG 299 (NE) | 3.93133 |
| Danio | ASP 84 (OD2) | ARG 299 (NH2) | 2.77819 |
| Danio | ASP 84 (OD2) | ARG 299 (NE) | 3.46174 |
| Danio | ASP 98 (OD1) | ARG 303 (NE) | 3.83540 |
| Danio | GLU 107 (OE1) | ARG 456 (NH1) | 2.28702 |
| Danio | GLU 107 (OE1) | ARG 456 (NE) | 2.19134 |
| Danio | GLU 107 (OE1) | ARG 456 (NH2) | 3.89534 |
| Danio | GLU 107 (OE2) | ARG 456 (NE) | 2.35397 |
| Danio | GLU 107 (OE2) | ARG 456 (NH2) | 2.67619 |

# Supplementary Table S2

Table 1: **Table S2.** Hydrogen bonds and salt bridges at the Mertk–Gas6 interface in Homo sapiens, Mus musculus, and Danio rerio complexes. Distances are reported in Å.

| Species | MERTK residue | Gas6 residue | Distance (Å) |
|---------|---------------|--------------|--------------|
| **Hydrogen bonds** | | | |
| Homo | LEU 270 (N) | SER 370 (OG) | 3.58711 |
| Homo | LEU 270 (N) | VAL 400 (O) | 2.98327 |
| Homo | VAL 272 (N) | LYS 402 (O) | 2.76290 |
| Homo | GLU 263 (OE1) | ARG 366 (NE) | 3.29144 |
| Homo | SER 273 (O) | ARG 366 (NH1) | 3.37414 |
| Homo | GLU 263 (OE2) | THR 368 (OG1) | 3.73627 |
| Homo | LEU 270 (O) | LYS 402 (N) | 3.01414 |
| Homo | VAL 272 (O) | ALA 404 (N) | 3.15327 |
| Mus | SER 230 (OG) | GLY 362 (O) | 2.07473 |
| Mus | LEU 265 (N) | SER 367 (OG) | 3.58003 |
| Mus | LEU 265 (N) | VAL 397 (O) | 3.04069 |
| Mus | VAL 267 (N) | LYS 399 (O) | 2.75868 |
| Mus | GLU 258 (OE1) | ARG 363 (NE) | 3.52161 |
| Mus | SER 268 (O) | ARG 363 (NH1) | 3.36457 |
| Mus | LEU 265 (O) | LYS 399 (N) | 3.02134 |
| Mus | VAL 267 (O) | ALA 401 (N) | 3.18877 |
| Danio | LEU 259 (N) | SER 359 (OG) | 3.56706 |
| Danio | SER 99 (OG) | GLU 387 (OE1) | 3.57946 |
| Danio | LEU 259 (N) | VAL 389 (O) | 2.98667 |
| Danio | ALA 261 (N) | LYS 391 (O) | 2.75846 |
| Danio | SER 262 (O) | ARG 355 (NH1) | 2.62597 |
| Danio | GLU 252 (OE1) | THR 357 (OG1) | 3.38783 |
| Danio | LEU 259 (O) | LYS 391 (N) | 2.97842 |
| Danio | ALA 261 (O) | ALA 393 (N) | 3.17327 |
| **Salt bridges** | | | |
| Homo | LYS 109 (NZ) | ASP 398 (OD1) | 3.97723 |
| Homo | LYS 109 (NZ) | ASP 398 (OD2) | 3.75752 |
| Homo | GLU 263 (OE1) | ARG 366 (NE) | 3.29144 |
| Homo | SER 273 (O) | ARG 366 (NH1) | 3.37414 |
| Homo | GLU 263 (OE1) | ARG 366 (NH2) | 3.25561 |
| Mus | GLU 258 (OE1) | ARG 363 (NE) | 3.52161 |
| Mus | SER 268 (O) | ARG 363 (NH1) | 3.36457 |
| Mus | GLU 258 (OE1) | ARG 363 (NH2) | 3.48912 |
| Danio | GLU 264 (OE1) | ARG 355 (NE) | 2.71740 |
| Danio | GLU 264 (OE1) | ARG 355 (NH1) | 3.62972 |
| Danio | GLU 264 (OE1) | ARG 355 (NH2) | 2.48863 |
| Danio | GLU 264 (OE2) | ARG 355 (NH2) | 3.36340 |

# Supplementary Table S3

Table 1: **Table S3.** Hydrogen bonds and salt bridges at the TYRO3–PROS1 interface in Homo sapiens, Mus musculus, and Danio rerio complexes.  Distances are reported in Å.

| Species | TYRO3  residue | PROTEIN  S  residue | Distance ( Å̃ ) |
|---------|----------------|---------------------|----------------|
| **Hydrogen  bonds** | | | |
| Homo | TYR 90 (N) | LYS 314 (O) | 2.73204 |
| Homo | GLN 88 (N) | ARG 316 (O) | 2.91951 |
| Homo | LYS 105 (NZ) | LYS 457 (O) | 3.21391 |
| Homo | LYS 61 (NZ) | SER 461 (O) | 2.28764 |
| Homo | GLU 71 (OE1) | TYR 312 (OH) | 3.87328 |
| Homo | TYR 90 (O) | LYS 314 (N) | 2.91296 |
| Homo | GLN 88 (O) | ARG 316 (N) | 2.95339 |
| Homo | ILE 74 (O) | ARG 316 (NH1) | 3.45717 |
| Homo | GLN 88 (OE1) | ARG 316 (NH1) | 3.76957 |
| Homo | TYR 90 (OH) | ARG 316 (NH1) | 3.25481 |
| Homo | ASP 73 (OD1) | ARG 316 (NH2) | 2.99023 |
| Homo | GLU 71 (OE1) | LYS 424 (NZ) | 3.06983 |
| Homo | ASP 87 (OD2) | GLN 458 (NE2) | 3.12408 |
| Mus | SER 82 (N) | TYR 312 (O) | 2.89395 |
| Mus | SER 80 (N) | LYS 314 (O) | 2.74523 |
| Mus | GLN 78 (N) | ARG 316 (O) | 2.89318 |
| Mus | LYS 95 (NZ) | LYS 457 (O) | 3.80712 |
| Mus | LYS 51 (NZ) | GLU 465 (OE1) | 3.00737 |
| Mus | LYS 51 (NZ) | GLU 465 (OE2) | 3.04718 |
| Mus | ASP 61 (OD1) | TYR 312 (OH) | 3.81878 |
| Mus | SER 80 (O) | LYS 314 (N) | 2.76957 |
| Mus | PRO 62 (O) | LYS 314 (NZ) | 3.51503 |
| Mus | GLN 78 (O) | ARG 316 (N) | 2.89630 |
| Danio | THR 70 (N) | VAL 311 (O) | 2.85553 |
| Danio | TYR 68 (N) | TYR 313 (O) | 2.76110 |
| Danio | GLN 66 (N) | ARG 315 (O) | 2.84662 |
| Danio | TYR 68 (O) | TYR 313 (N) | 2.95071 |
| Danio | GLU 49 (OE1) | TYR 313 (OH) | 2.52202 |
| Danio | GLN 66 (O) | ARG 315 (N) | 2.74201 |
| Danio | THR 64 (O) | ARG 317 (N) | 2.88696 |
| Danio | GLU 51 (OE1) | ARG 317 (NH2) | 2.64489 |
| Danio | GLU 49 (OE2) | TYR 426 (OH) | 3.79180 |
| **Salt  bridges** | | | |
| Homo | ASP 73 (OD1) | ARG 316 (NH2) | 2.99023 |
| Homo | GLU 71 (OE1) | LYS 424 (NZ) | 3.06983 |
| Mus | LYS 51 (NZ) | GLU 465 (OE1) | 3.00737 |
| Mus | LYS 51 (NZ) | GLU 465 (OE2) | 3.04718 |

| Species | TYRO3 residue | PROTEIN S residue | Distance ( Å ) |
|---|---|---|---|
| Mus | ASP 61 (OD2) | LYS 314 (NZ) | 3.25740 |
| Danio | ARG 39 (NE) | GLU 468 (OE1) | 3.09000 |
| Danio | ARG 39 (NH2) | GLU 468 (OE1) | 3.08000 |
| Danio | ARG 39 (NE) | GLU 468 (OE2) | 2.41000 |
| Danio | ARG 39 (NH1) | GLU 468 (OE2) | 2.27000 |
| Danio | GLU 51 (OE2) | ARG 315 (NH2) | 3.35000 |
| Danio | GLU 51 (OE1) | ARG 315 (NH2) | 2.55000 |
| Danio | GLU 51 (OE1) | ARG 317 (NE) | 3.61000 |
| Danio | GLU 51 (OE1) | ARG 317 (NH2) | 2.64000 |
| Danio | GLU 49 (OE2) | LYS 424 (NZ) | 2.99000 |
| Danio | GLU 49 (OE1) | LYS 424 (NZ) | 2.63000 |

Supplementary Table S4: AlphaFold-Multimer confidence and interface contact metrics for modeled TAM–ligand RBD complexes.

| Complex | ipTM | pTM | Chain A mean pLDDT | Chain B mean pLDDT | Interface res. A | Interface res. B | Contacts (4.5 Å) | Interface pLDDT (A) | Interface pLDDT (B) | Mean PAE (Å) |
|---|---|---|---|---|---|---|---|---|---|---|
| Axl (A)–Gas6 (B) (*Homo sapiens*) | 0.83 | 0.82 | 87.9 ± 13.0 | 92.3 ± 7.0 | 22 | 26 | 295 | 92.2 ± 5.2 | 86.5 ± 10.0 | 3.49 |
| Axl (A)–Gas6 (B) (*Mus musculus*) | 0.80 | 0.80 | 87.2 ± 14.0 | 90.7 ± 9.9 | 23 | 28 | 334 | 86.6 ± 13.9 | 81.6 ± 17.4 | 5.31 |
| Axl (A)–Gas6 (B) (*Danio rerio*) | 0.80 | 0.83 | 90.0 ± 6.7 | 89.4 ± 9.1 | 20 | 24 | 306 | 87.8 ± 7.6 | 85.5 ± 8.5 | 4.42 |
| Mertk (A)–Gas6 (B) (*Homo sapiens*) | 0.72 | 0.80 | 88.2 ± 11.7 | 91.0 ± 7.6 | 15 | 14 | 198 | 92.8 ± 4.0 | 92.3 ± 3.0 | 4.34 |
| Mertk (A)–Gas6 (B) (*Mus musculus*) | 0.78 | 0.82 | 88.5 ± 12.3 | 91.0 ± 7.8 | 15 | 16 | 185 | 94.1 ± 3.8 | 93.3 ± 2.7 | 3.11 |
| Mertk (A)–Gas6 (B) (*Danio rerio*) | 0.82 | 0.83 | 86.5 ± 10.8 | 87.2 ± 9.7 | 14 | 16 | 192 | 93.4 ± 3.4 | 89.5 ± 7.3 | 2.62 |
| Tyro3 (A)–Protein S (B) (*Homo sapiens*) | 0.77 | 0.80 | 90.6 ± 7.5 | 90.4 ± 7.9 | 20 | 17 | 229 | 87.6 ± 3.4 | 84.3 ± 11.5 | 4.11 |
| Tyro3 (A)–Protein S (B) (*Mus musculus*) | 0.79 | 0.81 | 91.9 ± 5.5 | 89.8 ± 8.7 | 20 | 17 | 222 | 88.9 ± 4.4 | 86.8 ± 6.5 | 3.52 |
| Tyro3 (A)–Protein S (B) (*Danio rerio*) | 0.84 | 0.81 | 93.5 ± 5.0 | 89.6 ± 9.9 | 18 | 19 | 256 | 89.9 ± 3.0 | 82.9 ± 11.4 | 3.57 |